\documentstyle[prd,aps,epsf,multicol]{revtex}
\begin{document}
\draft

\newcommand{\lsim}   {\mathrel{\mathop{\kern 0pt \rlap
  {\raise.2ex\hbox{$<$}}}
  \lower.9ex\hbox{\kern-.190em $\sim$}}}
\newcommand{\gsim}   {\mathrel{\mathop{\kern 0pt \rlap
  {\raise.2ex\hbox{$>$}}}
  \lower.9ex\hbox{\kern-.190em $\sim$}}}
\def\be{\begin{equation}}
\def\ee{\end{equation}}
\def\ba{\begin{eqnarray}}
\def\ea{\end{eqnarray}}
\title{Limiting SUSY-QCD spectrum and its application for decays of 
superheavy particles}

\author{V. Berezinsky$^{1,2}$ and M. Kachelrie{\ss}$^{1}$}

\address{$^1$INFN, Laboratori Nazionali del Gran Sasso,
         I--67010 Assergi (AQ), Italy \\
         $^2$Institute for Nuclear Research, Moscow, Russia}

\maketitle

\begin{abstract}
The supersymmetric generalization of the limiting and Gaussian QCD spectra is 
obtained. These spectra are valid for $x \ll 1$, when the main contribution 
to the parton cascade is given by gluons and gluinos. The derived 
spectra are applied to decaying superheavy particles with masses up
to the GUT 
scale. These particles can be relics from the Big Bang or produced by 
topological defects and could
give rise to the observed ultrahigh energy cosmic rays. 
General formulae for the fluxes of protons, photons and neutrinos 
due to decays of superheavy particles are obtained.
\end{abstract}



\section{Introduction}
The spectra of hadrons produced in deep-inelastic scattering and 
$e^+e^-$ annihilation are formed due to QCD cascading of the partons. 
In the Leading Logarithmic Approximation (LLA) which takes into
account $\ln(Q^2)$ terms this cascade is described by the 
Gribov-Lipatov-Altarelli-Parisi-Dokshitzer (GLAPD) equation \cite{GLAPD}.
The Modified Leading Logarithmic Approximation (MLLA) takes into
account additionally $\ln(x)$ terms, where 
$x=k_{\parallel}/k_{\parallel}^{max}$ and $k_{\parallel}$ is the  
longitudinal momentum of the produced hadron. Color coherence effects 
are described in MLLA. 
Two approximate analytic solutions to the MLLA evolution equations
have been obtained. 
These are the {\em limiting
spectrum\/} \cite{MLLA} and the {\em Gaussian spectrum\/} \cite{Mu,DFK},
in which we
include the {\em distorted Gaussian spectrum\/} \cite{distG,dkmt,esw}. 
The limiting spectrum is the most 
accurate one among them. In fact, it describes well the experimental data 
at large $x$, too, and this is natural, though accidental 
(for an explanation see 
Ref.~\cite{dkmt}). The limiting spectrum has a free normalization 
constant, $K_{\rm lim}$, which cannot be calculated theoretically and is found 
from comparison with experimental data. 
This constant has to be 
considered as a basic parameter of the theory, and it can be used at all 
energies, where the physical assumptions, under which the limiting spectrum 
is derived, are valid.  For detailed calculations of hadron spectra 
in $e^+e^-$ annihilation and comparison with experimental data see 
Ref.~\cite{comparison}. Up to energies of existing 
$e^+e^-$ colliders, $\sqrt{s} \lsim 183$~GeV, the limiting spectrum and the
distorted Gaussian spectrum describe well the available data. At large 
energies $\sqrt{s}\gsim 1$~TeV the production of supersymmetric 
particles might substantially change the QCD spectra. Apart from future 
experiments at LHC, supersymmetry (SUSY) might strongly reveal itself 
in the decays of superheavy particles. They can appear as relics of the  
Big Bang, or be produced by topological defects (TD), and can be the 
sources of the observed ultrahigh energy cosmic rays (UHECR) at 
$E \gsim 1\cdot 10^{10}$~GeV. The range of masses, $m_X$, of interest for 
UHECR goes from the GUT scale ($m_X \sim 10^{16}$~GeV) or less down to 
$m_X \sim 10^{12} - 10^{14}$~GeV.

In this paper we obtain the generalization  of the limiting and Gaussian 
QCD spectra for the SUSY case. Although the influence of SUSY on, e.g., the  
evolution of parton distributions or the running coupling constant
was considered in many works in the 80's \cite{split}, this is, to the best
of our knowledge, the first time that fragmentation spectra of hadrons
are examined for large $\sqrt{s}$ up to the GUT scale in SUSY-QCD. 
As application we use these spectra 
for calculations of the fluxes of UHECR produced in the decays of
superheavy particles.

\section{Limiting spectrum in SUSY-QCD}

The GLAPD equation \cite{GLAPD} describes in LLA the evolution of the 
parton distributions $D_A^B(x,\xi)$ with $\xi$.
Here $D_A^B$ is the distribution of partons $B$ inside the parton $A$
dressed by QCD interactions with coupling constant $\alpha(k_\perp^2)$, 
where $k_\perp$ is the transverse momentum and 
$x=k_{\parallel}/k_{\parallel}^{\rm max}$  is the longitudinal
momentum fraction of the parton $B$. The variable $\xi$
characterizes the maximum value of $k_\perp^2$ available in the
considered process ($k_\perp^2 < Q^2$),
\be
 \xi(Q^2)=\int_{\Lambda^2}^{Q^2} \frac{dk_\perp^2}{k_\perp^2} \:
                                 \frac{\alpha_s(k_\perp^2)}{4\pi} 
         \approx 
         \frac{\alpha_s(Q^2)}{4\pi}\,\ln\left(\frac{Q^2}{\Lambda^2}\right)\:,
\ee
with $\Lambda\sim 0.25$~GeV as phenomenological parameter.

In LLA, when terms with
$\alpha_s(Q^2) \ln(Q^2)$ are kept and terms proportional to $\alpha(Q^2)$ 
are neglected, the GLAPD equation can be written as \cite[Eq.~(1.79)]{dkmt}
\be   \label{GLAP}
 \frac{\partial}{\partial\xi} D_A^B(x,\xi)=
 \sum_C\int_0^1 \frac{dz}{z} \: \Phi_A^C(z) D_C^B(x/z,\xi) -
 \sum_C\int_0^1 dz \: z \Phi_A^C(z) D_A^B(x,\xi) \,,
\ee
where $\Phi_A^B(z)$ is the splitting function characterizing the
decay $A\to B+C$.

The supersymmetrization of Eq.~(\ref{GLAP}) is simple: each 
parton $A$ should be substituted by the supermultiplet which contains
$A$ and its superpartner $\tilde A$.
We shall generalize here the limiting spectrum of QCD given e.g. in \cite{dkmt}
to SUSY-QCD.
This spectrum is valid for small $x$, where gluons strongly dominate
the other partons. Therefore, we restrict ourselves to calculations
taking into account only two partons, namely gluons $g$ and gluinos
$\lambda$, in the tree diagrams. However, in the loop diagrams which
govern the running of $\alpha_s(k_\perp^2)$ we take into account also 
quarks and squarks.

Multiplying Eq.~(\ref{GLAP}) by $x^{j+1}$ and integrating it over
$x$, we obtain an equation for the moments $D_A^B(j,\xi)$,
\be   \label{GLAP,j}
 \frac{\partial}{\partial\xi} D_A^B(j,\xi)=
 \sum_C \Phi_A^C(z) D_C^B(j,\xi) -
 D_A^B(j,\xi) \sum_C\int_0^1 dz \: z\Phi_A^C(z)  \:,
\ee
where
\be
 D_A^B(j,\xi) = \int_0^1 dx \: x^{j-1}  D_A^B(x,\xi)
\ee
and the indices $A,B,C$ run through $g$ and $\lambda$.

The splitting functions 
are given e.g. in \cite{split} as
\ba
\Phi_g^g(z) &=& 4N_c \: \left[ \frac{1-z}{z}+\frac{z}{1-z}+z(1-z) \right] \\ 
\Phi_g^\lambda(z) &=& 2N_c \: \left[z^2+(1-z)^2\right] \\ 
\Phi_\lambda^g(z) &=& 2N_c \: \frac{1+(1-z)^2}{z} \\ 
\Phi_\lambda^\lambda(z) &=& 2N_c \: \frac{1+z^2}{1-z}  \:.
\ea

Equation~(\ref{GLAP,j}) can be rewritten in matrix form choosing as
basis $(g,\lambda)$,
\be    \label{S}
 \frac{\partial}{\partial\xi} D(j,\xi)= H(j)D(j,\xi)
\ee
where
\be   
 H(j) = \left( \begin{array}{cc}
               \nu_g(j) & \Phi_g^\lambda(j) \\
               \Phi^g_\lambda(j) & \nu_\lambda(j)
               \end{array}
        \right)
\ee
\be
 \nu_g(j) = \int_0^1 dz \left[ \left( z^{j-1}-z \right) \Phi_g^g(z)  
                               - \Phi_g^\lambda(z) \right]
\ee
\be      
 \nu_\lambda(j) = \int_0^1 dz \left( z^{j-1}-1 \right)
 \Phi_\lambda^\lambda(z) \:.
\ee
The procedure of solving Eq.~(\ref{S}) is identical to that given in
Ref.~\cite{dkmt}. After diagonalization of $H(j)$ with the help of 
$(D^+,D^-)$, the eigenvalues of $H$ are
\be
 \nu_\pm = \frac{1}{2} \left( \nu_g + \nu_\lambda \pm
           \left[ \left( \nu_g - \nu_\lambda \right)^2
                  + 4 \Phi_g^\lambda\Phi^g_\lambda \right]^{1/2}
\right) \:.
\ee
In the limit $\omega=j-1\to 0$, the leading term $\nu_+$ is  given by
\be
 \nu_+ = \frac{4N_c}{\omega} -a + O(\omega)
\ee
with $a=\frac{11}{3} N_c=11.$ ($N_c=3$ is the number of colours.)


Up to now we have considered the LLA approximation. This
approximation is not correct for $x\ll 1$, when colour coherence
effects become important. A better description of distributions at small
$x$ is given by the MLLA approximation \cite{MLLA}. It takes into account
both $\ln(Q^2)$ and $\ln(x)$ terms as well as angular ordering.
The MLLA evolution equation results in Eq.~(\ref{S}) for the moments,
with $\xi$ replaced by 
$\xi_{\rm MLLA} = \alpha_s(Y)/(4\pi) \,\ln(Y^2)$ with
$Y=\ln(E\theta/Q_0)$, where $E$ is the energy
and $\theta$ the opening angle of the jet (see \cite[Eq.~(7.4)]{dkmt}.
For its analytic solution at small $x$, the limiting spectrum, the
eigenvalues (13) derived above in LLA are still valid.
Properly normalized, the limiting spectrum $D_{\rm lim}(l,Y)=
xD_{\rm lim}(x,Y)$ gives
$\sigma^{-1}d\sigma/dl$ in the case of $e^+e^-$ annihilation and the
decay spectrum of $X$ particles, where $l=\ln(1/x)$.
This spectrum is given by \cite{kh96}

%
%
%
%
\be
 D_{\rm lim}(l,Y) = K_{\rm lim} \,\frac{4C_F}{b}\,\Gamma(B)
      \int_{-\pi/2}^{\pi/2} \frac{d\tau}{\pi}\: e^{-B\alpha} 
      \left( 
        \frac{b}{8N_c}\:\frac{\sinh\alpha}{\alpha}\:\frac{y}{Y} 
      \right)^B  I_B (y)  \:.
\ee
Here, $Y$ now is $Y=\ln[\sqrt{s}/(2\Lambda)]$, 
$\alpha=\alpha_0 +{\rm i}\tau$, $\alpha_0={\rm arctanh}(2\zeta-1)$,
$\zeta=1-l/Y$,  and $\sqrt{s}$ is the
c.m. energy of an $e^+e^-$ pair or the mass $m_X$ of the superheavy
decaying particle. 
The parameters depending on the structure of the
theory are $a=\frac{11}{3} N_c$, the constant $b$ of evolution of
$\alpha_s(k_\perp^2)$ in one-loop approximation,
$k_\perp^2 d\alpha_s(k_\perp^2)/dk_\perp^2=-b\alpha_s^2/(4\pi)$, and 
$C_F=(N_c^2-1)/(2N_c)=4/3$. 
When the masses of the superheavy coloured Higgses are larger than
$Q^2=k_{\perp,{\rm max}}^2$, $b=b_{\rm SUSY}=9-n_f$, 
where $n_f=6$ is the number of quark flavours. Finally,
$I_B$ is the modified Bessel function of order $B=a/b$,  and argument
\be
 y(\tau)=\left( \frac{16N_C}{b}\frac{\alpha}{\sinh\alpha}\:
                [\cosh\alpha+(1-2\zeta)\sinh\alpha] \: Y \right)^{1/2} \:.
\ee
A convenient way for the numerical evaluation of $I_B(y)$ is the use of
its series expansions, given for example by (8.445) and (8.451.5) in 
Ref.~\cite{gr}.

In Fig.~1 the SUSY-QCD limiting spectrum is shown in comparison with 
the QCD limiting spectrum, for the case of three colours, $N_c=3$, and
six quarks flavours, $n_f=6$. For the sake of comparison, we normalize
both spectra by the condition  
\be     \label{norm}
 \int_0^1 dx \: xD_{\rm lim} (x,Y) = 2  \:.
\ee
The maximum of the spectra is at $l_m=Y(0.5+\sqrt{c/Y}-c/Y)$ with
$c\approx 0.39$ in QCD and $c\approx 0.84$ in SUSY-QCD. Therefore, the SUSY 
spectra are shifted to the right, and since they are also narrower than the
QCD spectra (see Eq.~(\ref{sigma}) below), the SUSY maxima are dramatically
higher (by a factor of 30) than the QCD ones. We remind the reader that 
the value of the maximum is given by the multiplicity.

We compare these spectra also with the Gaussian approximation obtained
for the MLLA solution \cite{distG,esw}. This approximation can be easily
generalized to the SUSY case \cite{bmv} and has for $x\ll 1$ as function
of $x=2E/\sqrt{s}$ the form,
\be \label{gauss}
 D_G (x,Y)=\frac{K_G}{x} \: \exp\left( -\frac{\ln^2 x/x_m}{2\sigma^2}
                                \right)
\ee
Assuming one-loop SUSY evolution of $\alpha_s(k_\perp^2)$ with 
$b=b_{\rm SUSY}$ and $\Lambda=\Lambda_{\rm QCD}$, one has
\be  \label{sigma}
 \sigma^2 = \frac{1}{24}\:\sqrt{\frac{b_{\rm SUSY}}{6}} \: 
            \ln^{3/2} \left( \frac{s}{\Lambda^2} \right) 
\ee
and
\be
 x_m = \left( \frac{\Lambda}{\sqrt{s}} \right)^{1/2} \:.
\ee
Since $b_{\rm SUSY}=3$ (for $n_f=6$) is less than $b_{\rm QCD}=7$ (for
the same $n_f$), the SUSY-QCD peak is narrower than the QCD one.

The Gaussian spectrum given by Eq.~(\ref{gauss}) has its maximum at $x=x_m$.
It gives a less precise description than the
limiting spectrum. One might expect that at large $Y$, when the
higher momenta, skewness $s$ and kurtosis $\kappa$, become small
($s\sim Y^{-3/2}$ and $\kappa\sim Y^{-1/2}$) the agreement improves.
We have found that 
both for the case of ordinary QCD and SUSY-QCD the shape of the
spectra differs substantially in the interesting range $10^{-6} \lsim x
\lsim 10^{-2}$ (Fig.~2).

\section{Applications}

We shall apply now the limiting spectrum in SUSY-QCD to the
calculation of the spectrum of ultrahigh energy cosmic rays (UHECR)
generated by the decay of superheavy particles with masses
$10^{12}-10^{15}$~GeV.

Let us discuss first the problem of the normalization of spectrum.  
We remind the reader that the limiting spectrum has been derived for
$x\ll 1$, though at least for small $s$ it describes well the
experimental data for $x$ up to $x\sim 1$ (see \cite{dkmt} for
discussion). In Ref.~\cite{kh96}, the
normalization constant was fixed by a comparison with experimental data 
on $e^+e^-$ annihilation to $K_{\rm lim}=2.6$. This value is fixed by 
$D_{\rm lim}(l,Y)$ at maximum (see Fig.~1), i.e. by multiplicity. 

Since the shape of the
spectrum and the position of its peak change dramatically, if one goes
from QCD and $\sqrt{s}\sim 100$~GeV to SUSY-QCD and $\sqrt{s}\sim
10^{12}-10^{15}$~GeV, we cannot use this value of $K_{\rm lim}$.
Instead we use as normalization condition Eq.~(\ref{norm}) replacing
the factor 2 by $2f_i$,
where $i$ runs through $N$ (all nucleons) and $\pi^\pm$ and $\pi^0$
(charged and neutral pions), while $f_i$ is the fraction of energy
carried by the hadron $i$.
Note that the main contribution to the integral in Eq.~(\ref{norm})
comes from large values 
$x\sim 1$, where the limiting  spectrum might have large uncertainties.
However, this is, in our opinion, the most physical way of normalization.
The numerical values of $f_i$ are unknown at large $s$. One can assume
that $f_\pi\approx 1-f_{\rm LSP}$, where $f_{\rm LSP}$ is the energy fraction
taken away by the lightest supersymmetric particle (LSP). According to
the simplified Monte-Carlo simulation of \cite{bk98}, $f_{\rm LSP}\sim
0.4$. For the ratio $f_N/f_\pi$ we use $\sim 0.05$ inspired by $Z^0$ decay.

Let us assume that the decay rate of $X$ particles $\dot n_X$ in the
extragalactic space does not depend on distance and time. Then taking
into account the energy losses of UHE protons and the absorption of
UHE photons due to pair production ($\gamma+\gamma\to e^+ +e^-$) on the
radio and microwave background, the diffuse flux of UHE protons and
antiprotons is
\be
 I_{p+\bar p} (E) = \frac{1}{2\pi} \: \frac{\dot n_X}{m_X}
                    \int_0^\infty dt_g D_N(x_g,Y)\frac{dE_g(E,t_g)}{dE} \:,
\ee
where 
$E_g(E,t_g)$ is the energy at generation time $t_g$ of a proton which has
at present the energy  $E$ and $x_g= 2E_g /m_X$.
Denoting the proton energy looses on microwave radiation by  $dE/dt=b(E,z)$, 
$dE_g/dE$ is given by \cite{bbgdp} 
\be  \label{looses}
  \frac{dE_g(E,z_g)}{dE} = (1+z_g) \exp\left[ \int_0^{z_g} \frac{dz}{H_0}
  \: (1+z)^{1/2} \left( \frac{\partial b(E,0)}{\partial E} \right)_{E=E_g(z)}
  \right] \:,
\ee
where $H_0$ is the Hubble constant and $z$ the redshift.
In the case that the energy looses on microwave radiation are much larger
than the adiabatic ones, Eq.~(\ref{looses}) reduces to 
\be
  \frac{dE_g(E,z_g)}{dE} \approx \frac{b(E_g,0)}{b(E,0)} \:.
\ee

The diffuse spectrum of UHE photons can be calculated as
\be 
 I_\gamma(E) = \frac{\dot n_X}{\pi} \: \lambda_\gamma(E) \:
               \frac{1}{m_X} \int_{2E/m_x}^1 \frac{dx}{x} \: 
                D_{\pi^0} (x,Y) \:,
\ee
where $\lambda_\gamma(E)$ is the absorption length of a photon.
For the numerical evaluation of $I_\gamma$, we use $\lambda_\gamma$
from \cite{pb96}.

The neutrino flux depends generally on the evolution of the sources,
\be  \label{n_X(t)}
 \dot n_X (t) = \dot n_X (t_0) \left(\frac{t_0}{t}\right)^{3+p} \:.
\ee  
The number of neutrinos with energy $E$ produced per decay of one $X$ 
particle can be approximately calculated as
\be   \label{N_nu}
  N_\nu(E) \approx \frac{12}{m_X} \int_{4E/m_x}^1 \frac{dx}{x} \: 
                D_{\pi^\pm} (x,Y) \:.
\ee
Combining Eqs.~(\ref{n_X(t)}) and (\ref{N_nu}), we obtain for the diffuse
neutrino flux
\be    \label{nu}
 I_\nu(E) = \frac{3\dot n_X(t_0)}{\pi H_0 m_X} 
            \int_0^{z_{\rm max}} dz \: (1+z)^{\frac{3}{2}p}  
            \int_{\frac{4E(1+z)}{m_x}}^1 \frac{dx}{x} \: D_{\pi^\pm} (x,Y) \:, 
\ee
where $1+z_{\rm max}(E)\approx m_X/(4E)$.

In Fig.~3, the spectra of UHE protons, photons, and neutrinos are
shown together with experimental data \cite{na}
for the model of cosmic necklaces -- topological defects which
consists of monopoles connected by strings \cite{hk}. We use the model of 
Ref.~\cite{bv}, where
\be  \label{neck}
 \dot n_X = \frac{r^2\mu}{m_X t^3}
\ee
with $r^2\mu=5\cdot 10^{27}$~GeV$^2$ and $m_X=1\cdot 10^{14}$~GeV.
The proton flux is suppressed at the highest energies as compared with
the calculations of Ref.~\cite{bv}, where the Gaussian SUSY-QCD spectrum
was used. The model corresponds to $p=0$, or effectively to the absence of
evolution  for the integral over $z$ in Eq.~(\ref{nu}).

In conclusion, we have calculated the SUSY-QCD limiting spectrum of
partons in a 
jet. This spectrum considerably differs from that of ordinary QCD: the 
maximum of the Gaussian peak is shifted towards smaller $x$ and the peak is 
narrower and higher. The limiting spectrum has been applied for calculations
of the UHE fluxes
of protons, photons and neutrinos from decays of superheavy particles in 
the Universe. The fluxes of these particles are different 
from those calculated with the ordinary QCD parton spectra.

\acknowledgments
We are deeply grateful to Yu.L. Dokshitzer for many useful explanations and 
comments and to V.A. Khoze for stimulating conversations. S.I. Grigorieva 
provided us with her new calculations (unpublished) of proton energy losses 
on microwave radiation and of the derivative $\partial b(E,0)/\partial E$.
We are grateful to M. Birkel for interesting and useful
correspendence, to P. Blasi and A. Vilenkin for permanent discussions
and to an anonymous referee for helpful and clarifying remarks.
MK would like to thank the Alexander von Humboldt-Stiftung for 
a Feodor-Lynen grant.


\end{document}